\documentclass[aps,prl,twocolumn,superscriptaddress,floats,showpacs]{revtex4}
\usepackage{}
\usepackage{txfonts}
\usepackage{amssymb}
\usepackage{graphicx}

\begin{document}

\title{An ARPES Study of the Electronic Structure of the Quantum Spin Liquid EtMe$_3$Sb[Pd(dmit)$_2$]$_2$}

\author{Q. Q. Ge}

\author{H. C. Xu}

\author{X. P. Shen}

\author{M. Xia}

\author{B. P. Xie}

\author{F. Chen}

\author{Y. Zhang}

\affiliation{State Key Laboratory of Surface Physics, Department of Physics,  and Advanced Materials Laboratory, Fudan University, Shanghai 200433, People's Republic of China}

\author{R. Kato}

\affiliation{Condensed Molecular Materials Laboratory, RIKEN, Wako-shi, Saitama 351-0198, Japan}

\author{T. Tsumuraya}

\affiliation{Condensed Molecular Materials Laboratory, RIKEN, Wako-shi, Saitama 351-0198, Japan}

\affiliation{Computational Materials Science Unit, National Institute for Materials Science, Tsukuba, Ibaraki 305-0047, Japan}

\author{T. Miyazaki}

\affiliation{Computational Materials Science Unit, National Institute for Materials Science, Tsukuba, Ibaraki 305-0047, Japan}

\author{M. Matsunami}

\author{S. Kimura}

\affiliation{UVSOR Facility, Institute for Molecular Science and The Graduate University for Advanced Studies, Okazaki 444-8585, Japan}

\author{D. L. Feng}\email{dlfeng@fudan.edu.cn}

\affiliation{State Key Laboratory of Surface Physics, Department of Physics,  and Advanced Materials Laboratory, Fudan University, Shanghai 200433, People's Republic of China}

\date{\today}

\begin{abstract}

The electronic structure of  a quantum spin liquid compound, EtMe$_3$Sb[Pd(dmit)$_2$]$_2$, has been studied with angle-resolved photoemission spectroscopy, together with two other Pd(dmit)$_2$ salts in the valence bond solid or antiferromagnetic state. We have resolved several bands that have  negligible dispersions and fit well to the calculated energy levels of an isolated [Pd(dmit)$_2$]$_2$ dimer.  EtMe$_3$Sb[Pd(dmit)$_2$]$_2$  being a  Mott insulator,  its  lower Hubbard band  is identified, and there is a small gap of $\sim$50~meV between this band and the chemical potential.  Moreover, the spectral features exhibit polaronic behavior with anomalously broad linewidth. Compared with existing theories, our results suggest that strong electron-boson interactions, together with smaller hopping and on-site Coulomb interaction terms have to be considered for a realistic modeling of the organic quantum spin liquid systems like the Pd(dmit)$_2$ salt.
\end{abstract}

\pacs{71.36.+c, 79.60.Fr, 71.20.Rv}

\maketitle


Quantum spin liquid (QSL) is an exotic state of quantum matter, which  has profound implications for magnetism and high temperature superconductivity \cite{PALeeRMP}.  For  two-dimensional (2D)  spin-1/2 systems in the presence of antiferromagnetic (AFM) interactions, QSL might be realized if the geometrical frustration is so strong that it completely destroys the long-range magnetic order even at zero temperature, giving rise to a large degeneracy of the ground states.  Although the QSL state on a 2D triangular lattice was first proposed in 1973 \cite{firsttheory}, it is hard to achieve experimentally, because the   finite three-dimensional (3D) magnetic interactions often help drive the system into certain ordered states. Recently, some promising QSL candidates have been reported \cite{kCN,dmit,CuBDC,ZnCu,BaCu,NaIrO,Spinice,CNMR2,H3QSL,Kag}, for example, $\kappa$-(BEDT-TTF)$_2$Cu$_2$(CN)$_3$ \cite{kCN} and EtMe$_3$Sb[Pd(dmit)$_2$]$_2$ \cite{dmit} on 2D triangular lattices, BaCu$_3$V$_2$O$_8$(OH)$_2$ \cite{BaCu}  on a Kagome lattice, Na$_4$Ir$_3$O$_8$ \cite{NaIrO} on a hyper-Kagome lattice,  and Ho$_2$Ti$_2$O$_7$ on a pyrochlore lattice \cite{Spinice}. They all show no sign of long-range order down to very low temperatures.


\begin{figure}[t!]
\includegraphics[width=8.6cm]{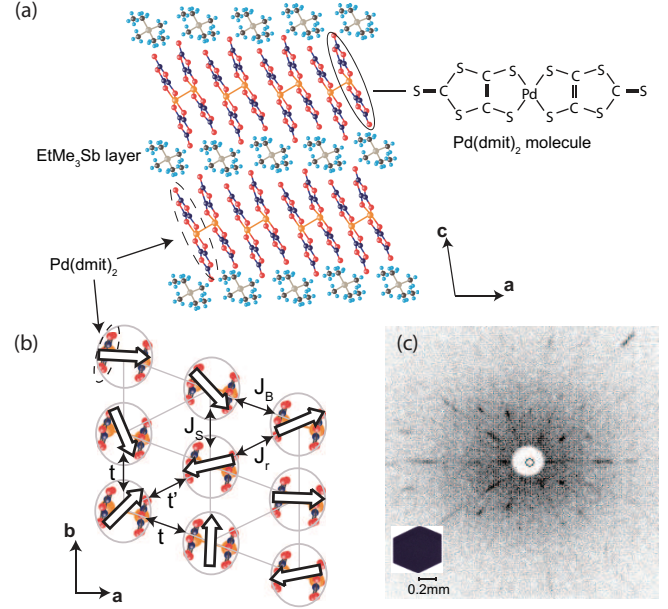}\caption{(color online)
The  structure  of EtMe$_3$Sb[Pd(dmit)$_2$]$_2$. (a) A view from the \textit{\textbf{b}} axis. The crystal structure is monoclinic, and consists of two-dimensional Pd(dmit)$_2$ layers separated by layers of cation EtMe$_3$Sb. (b) An in-plane sketch of the spin arrangement on the Pd(dmit)$_2$ layer. Each lattice (encircled by solid lines) stands for strongly dimerized [Pd(dmit)$_2$]$_2^{-}$. Three nearest-neighbor exchange interactions, \emph{J$_B$},\emph{ J$_S$} and \emph{J$_r$}, are similar and estimated to be 220~--~250~K. (c) The Laue picture of the EtMe$_3$Sb[Pd(dmit)$_2$]$_2$ single crystal. Because the dimer units  in panel~a stack along the \textit{\textbf{a}}+\textit{\textbf{b}} direction in one layer and along the \textit{\textbf{a}}-\textit{\textbf{b}} direction in the next layer \cite{crystals}, the crystal shows twofold symmetry. The inset shows the photo of a single crystal.}
\label{sample}
\end{figure}

EtMe$_3$Sb[Pd(dmit)$_2$]$_2$ (dmit=1,3-dithiole-2-thione-4,5-dithiolate, Me=CH$_3$, Et=C$_2$H$_5$), as shown in Fig.~\ref{sample}(a), provides a physical realization of a frustrated spin-1/2 QSL system on a triangular lattice \cite{dmit, CNMR, ThermoCon}. In fact, a series of layered organic salts X[Pd(dmit)$_2$]$_2$ are Mott insulators at ambient pressure \cite{crystals}, where X is a nonmagnetic monovalent cation such as Et$_2$Me$_2$P, EtMe$_3$Sb, EtMe$_3$P, \emph{etc}. The Pd(dmit)$_2$ units are strongly dimerized, forming a spin-1/2 unit [Pd(dmit)$_2$]$_2^{-}$ [Fig.~\ref{sample}(b)].  The recent first-principles calculations on a series Pd(dmit)$_2$ salts  have shown that the  choice of X would affect the hopping integrals ($t$, $t'$), and the strength of the frustration  ($J'/J$) \cite{Theory2,arXivCalc},  where  the exchange interaction terms \emph{$J'$ = J$_r$ }and \emph{J = J$_B$ $\simeq$ J$_S$} [Fig.~\ref{sample}(b)]. On the other hand, the effective on-site Coulomb interaction ($U$) is weakly dependent on the choice of X \cite{crystals}.  The different parameters would then give  different ground states \cite{arXivCalc,Theory2}. For example, besides the QSL, Et$_2$Me$_2$P[Pd(dmit)$_2$]$_2$ exhibits AFM order below $\thicksim$~20~K \cite{Trans1},  EtMe$_3$P[Pd(dmit)$_2$]$_2$ is a valence bond solid or VBS, where two neighboring sites form a spin singlet without magnetic order \cite{VBS},  and Et$_2$Me$_2$Sb[Pd(dmit)$_2$]$_2$ is in a charge ordered state  \cite{COT}.

All the QSL, VBS, and AFM Pd(dmit)$_2$ salts are Mott insulators, so they are often  modelled using the Hubbard model.
The  band calculations give metallic ground states with sizable  \emph{t} \cite{Theory2, arXivCalc, CalcAFM, calc3}. Then like for the cuprate, $U$, or the ratio of $U/t$, is critical in determining different ground states here \cite{Theory1, Theory3}. It has been argued that by reducing $U/t$, the system goes from a phase with long-range spiral magnetic order, to a spin liquid state, and then to a metallic phase \cite{Theory1}. That is, the spin liquid state on a triangular lattice is stable over a narrow range on the verge of metal-insulator transition, suggesting that the electron itinerancy may play a role in realizing spin liquid state \cite{Theory1, Theory3, annurev}. More specifically,  for the   EtMe$_3$Sb[Pd(dmit)$_2$]$_2$ QSL compound, its \emph{t} was estimated to be $\sim$ 30-55~meV \cite{Theory2,arXivCalc,calc3}, and \emph{U} has been estimated to be about 0.7~eV \cite{calc3}. However, experimentally,   little  is known  on  the basic electronic structure of a QSL, such as its  band structure and  Mott gap (or $U$), and whether the Hubbard model is sufficient.

In this paper, we report the detailed electronic structure of EtMe$_3$Sb[Pd(dmit)$_2$]$_2$ measured with angle-resolved photoemission spectroscopy (ARPES). We find three broad spectral features with negligible dispersion within the first 1.5~eV binding energy range, which match well with the calculated energy levels of the isolated dimer. The weak dispersion and broad lineshape resemble the electronic structure in polaronic systems, indicating strong interactions between electrons and certain bosonic modes in the system. The feature near the chemical potential ($\mu$) is recognized as the lower Hubbard band (LHB), and the Mott insulator character is evidenced by the lack of spectral weight within  $\sim$50~meV of the chemical potential. QSL compound is found to be more insulating than  VBS and AFM Pd(dmit)$_2$ salts. Our results  indicate that the experimental  $t$ and $U$ terms of EtMe$_3$Sb[Pd(dmit)$_2$]$_2$ are much smaller than what were previously considered,  which naturally explain its various anomalous  properties.  We suggest that the  Hubbard model is not sufficient for describing  EtMe$_3$Sb[Pd(dmit)$_2$]$_2$  or the organic QSL systems in general, instead, a Holstein-Hubbard model with strong electron-phonon/magnon interactions is more appropriate.

The ARPES measurements were conducted at Beamline 7 of Ultraviolet Synchrotron Orbital Radiation Facility (UVSOR). The overall energy resolution was set to 20~meV or better and the typical angular resolution is 0.5$^\circ$. Fine single crystals of EtMe$_3$Sb[Pd(dmit)$_2$]$_2$, EtMe$_3$P[Pd(dmit)$_2$]$_2$, and Et$_2$Me$_2$P[Pd(dmit)$_2$]$_2$ were synthesized by an aerial oxidation method \cite{crystals}, pre-oriented by Laue diffraction, then cleaved \emph{in situ} and measured under ultrahigh vacuum of 6 $\times$ 10$^{-11}$ mbar. During measurements the temperature was kept at 70 K to avoid the onset of charging at 40 K. Most importantly, we employed low energy photons as the excitation light source and kept the samples exposed under low photon flux to reduce the radiation damage, so that the aging effects are negligible in the data.


 The single crystal of EtMe$_3$Sb[Pd(dmit)$_2$]$_2$ is usually a hexagonal piece with typical size of 0.6 $\times$ 0.6 $\times$ 0.05~mm$^3$ as shown in the inset of Fig.~\ref{sample}(c). The unit cell is monoclinic as shown in Fig.~\ref{sample}(a) and the space group is \emph{C2/c}. More detailed description of the crystal structure could be found in Ref.~\cite{crystals}. The high sample quality was confirmed by the typical Laue pattern [Fig.~\ref{sample}(c)], which manifests the twofold symmetry of the crystal structure.


\begin{figure}[t]
\includegraphics[width=8.6cm]{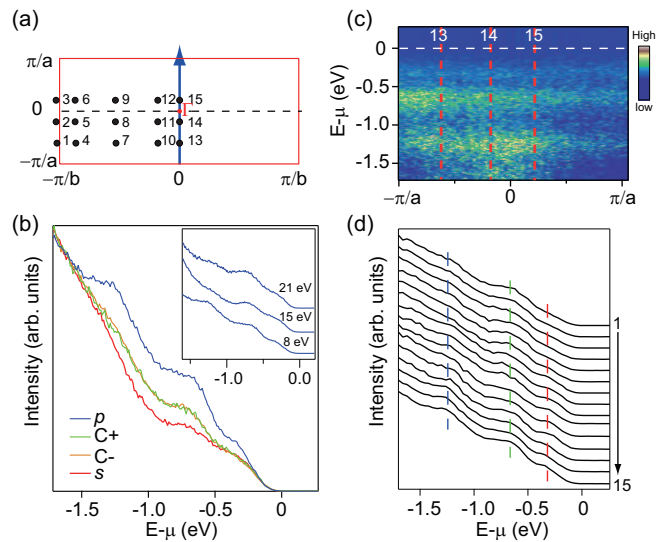}\caption{(color online)
Photoemission data of EtMe$_3$Sb[Pd(dmit)$_2$]$_2$. (a) The two-dimensional Brillouin zone of EtMe$_3$Sb[Pd(dmit)$_2$]$_2$. (b) The normal-emission photoemission spectra of EtMe$_3$Sb[Pd(dmit)$_2$]$_2$ excited by \emph{p}, \emph{s}, left-handed circular (C+) and right-handed circular (C-) polarized 8~eV photons. The inset shows the normal emission spectra collected with \emph{p} polarized 8, 15, 21~eV photons. (c) The photoemission intensity along the (0, -$\pi$/a)--(0, $\pi$/a) line as indicated in panel~a. Here in order to highlight the main spectral features, the original data has been deducted off the Shirley background obtained in Fig.~\ref{fitting}(a) to exclude the influence of secondary electrons. (d) ARPES spectra of EtMe$_3$Sb[Pd(dmit)$_2$]$_2$ acquired at  momenta  \#1~--~\#15 as coded in panel~a. The photoemission data in panels~c and d were taken with \emph{p} polarized 8~eV photons. }
\label{spectra}
\end{figure}

\begin{figure}[t]
\includegraphics[width=8.6cm]{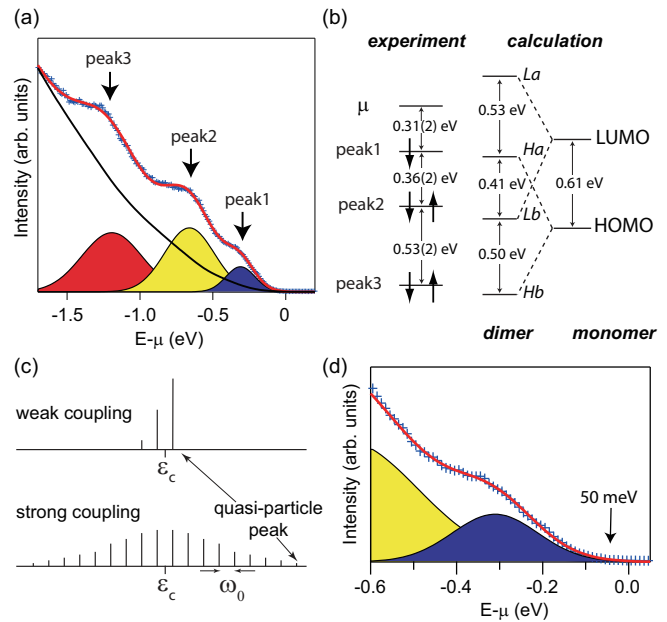}\caption{(color online)
(a) The normal-emission valence band spectra of EtMe$_3$Sb[Pd(dmit)$_2$]$_2$ is fitted with three Gaussians  plus a Shirley-type background. (b) Left: The centroid of the peaks obtained from the fitting. The black arrows represent the electron spins. Right: The calculated energy levels of the isolated Pd(dmit)$_2$ monomer and [Pd(dmit)$_2$]$_2$ dimer in EtMe$_3$Sb[Pd(dmit)$_2$]$_2$. The antibonding LUMO, antibonding HOMO, bonding LUMO, and bonding HOMO energy levels are labeled as \textit{La}, \textit{Ha}, \textit{Lb}, and \textit{Hb}, respectively. (c) A schematic spectral function of a single electron coupled to a bath of bosons of frequency $\omega_0$ in the weak coupling and strong coupling conditions. (d) The spectrum near the chemical potential is enlarged, and a gap of 50~meV is hereby defined. }
\label{fitting}
\end{figure}

The normal-emission spectra from the $\Gamma$-point of EtMe$_3$Sb[Pd(dmit)$_2$]$_2$  Brillouin zone [Fig.~\ref{spectra}(a)] are displayed in Fig.~\ref{spectra}(b). In order to make sure all states are detected, various photon polarizations were exploited, including the in-plane (\emph{p}), perpendicular-to-plane (\emph{s}), left-handed circular and right-handed circular polarizations. Here, the ``plane" is defined by the incident photon beam direction and the outgoing photoelectron trajectory, so that \textit{s} and \textit{p} polarizations are sensitive to only odd and even electron wavefunctions with respect to this mirror plane \cite{ZhangBaCo, ZhangNaFeAs}. We found three broad features within the 1.5~eV binding energy range, and they are most pronounced in the $p$-polarization data. Moreover, their binding energies are robust against the variation of the photon energy as illustrated in the inset of Fig.~\ref{spectra}(b), which is a manifestation of the 2D character of the system. Hereafter, we only show the data taken with \emph{p} polarized 8~eV photons, which give the most pronounced spectral features. Fig.~\ref{spectra}(c) illustrates the  photoemission intensity along (0, -$\pi$/a)--(0, $\pi$/a) after deducting a Shirley background as show in Fig.~\ref{fitting}(a), which shows spectral features with negligible dispersion.  Furthermore,  as displayed in Fig.~\ref{spectra}(d), all the photoemission spectra taken  at various momenta in the first Brillouin zone show three specific spectral features as marked by the dashed lines. However, no obvious momentum dependence of the peak positions is observed, indicating the states are considerably localized in EtMe$_3$Sb[Pd(dmit)$_2$]$_2$,  contradicting  the  pronounced dispersions up to 400~meV  predicted along (0, -$\pi$/a)--(0, $\pi$/a) and (-$\pi$/b, 0)--($\pi$/b, 0)  in the calculations \cite{Theory2, arXivCalc, calc3}.


The photoemission data of EtMe$_3$Sb[Pd(dmit)$_2$]$_2$ show no density of state (DOS) at the chemical potential [Fig.~\ref{fitting}(a)], as expected for a Mott insulator. Recent torque magnetometry measurement on EtMe$_3$Sb[Pd(dmit)$_2$]$_2$ revealed an extended quantum critical phase, in which low-lying gapless spinon excitations behave like the elementary excitations in paramagnetic metals with Fermi surface \cite{TorqueMag}.  We here do not observe any projection of the spinon Fermi surface in the charge channel. Furthermore, we found that the photoemission spectrum is best fitted with three Gaussian components, plus a Shirley-type background in order to account for secondary electrons [Fig.~\ref{fitting}(a)]. To compare with the experiment, we have  calculated the  energy levels of the isolated Pd(dmit)$_2$ monomer and [Pd(dmit)$_2$]$_2$ dimer in a supercell using their structures in EtMe$_3$Sb[Pd(dmit)$_2$]$_2$ with the all-electron full-potential linearized augmented plane wave (FLAPW) method. The resulting energy levels are almost the same as those calculated in Ref.~\cite{calc1} for another [Pd(dmit)$_2$]$_2$ salt. As illustrated in Fig.~\ref{fitting}(b), the interactions between the two Pd(dmit)$_2$ molecules in an isolated dimer will split their respective HOMO (highest occupied molecular orbital) and LUMO (lowest unoccupied molecular orbital) states into  bonding and  antibonding states, which are labeled as \textit{Hb}, \textit{Ha}, \textit{Lb} and \textit{La} self-explanatorily. Based on the stoichiometry, there are five electrons, resulting in a half-filled \textit{Ha} band. It is also the fully filled LHB of this Mott insulator.  Based on the fitting, the three Gaussian components located at 0.31~$\pm$~0.02, 0.67~$\pm$~0.02, 1.20~$\pm$~0.02~eV and labeled as peak1, peak2, and peak3, which  correspond to the \textit{Ha}/LHB, \textit{Lb}, and \textit{Hb} bands, respectively [Fig.~\ref{fitting}(b)]. The agreement between the experiment and calculation is reasonably good, especially considering that there are small difference between the intra-orbital and inter-orbital onsite Coulomb interactions \cite{calc3}. Moreover, the spectral weight of peak2 and peak3 are comparable, while peak1 is considerably weaker, which is expected as the LHB is only filled with one electron.


The localized electronic structure is not only manifested by the negligible dispersion, but also by the broad linewidth. The fitted full width at half-maximum (FWHM) of peak1, peak2 and peak3 are 0.23, 0.40, and 0.48~eV, respectively. In some calculations, there are a set of bands around 1.2~eV, which may cause the possible broad lineshape of peak3, but the spectral width of peak2 is much broader than the calculated bandwidth of 0.1~eV \cite{Theory2, calc3, arXivCalc}. The broad lineshape that could be fitted with the incoherent Gaussian peaks resembles those of polaronic systems, where they are caused by strong electron-phonon coupling \cite{shenNaTaS, Grioni, Mannella}. As shown in Fig.~\ref{fitting}(c), the quasiparticle weight is usually vanishingly small at the low-energy edge of the spectrum, and the broad spectrum is an envelope of many individual peaks due to strong electron-boson interactions \cite{polaron}. For a flexible organic compound with strong spin fluctuations, strong electron-phonon interactions and/or electron-magnon interactions are expected \cite{polaron3}. Therefore, the charge carriers in QSL could be regarded as polarons, which are heavily dressed by phonons (lattice deformations) or magnons. In addition, it has been shown  that the centroid of the incoherent spectral distribution would follow the bare band dispersion in the absence of strong correlations \cite{polaron2, BPXie}. Therefore, our data indicate that \emph{t}  is at most a few meV, much smaller than that predicted in the previous calculations  \cite{Theory2, arXivCalc, calc3}. It shows that dimers are rather isolated in the crystal, which is consistent with the fact that the centroid positions of the spectral features match the energy levels of the isolated dimer in Fig.~\ref{fitting}(b).

Taking the lower edge of the LHB band as the coherent quasiparticle location of the polaronic lineshape [Fig.~\ref{fitting}(c)], a gap could be estimated as in Fig.~\ref{fitting}(d). It is about 50~meV, or smaller, because the DOS gradually decreases to zero at the chemical potential. If QSL appears only in a narrow region of the \emph{U/t}  axis  \cite{Theory1, Theory3}, the small hopping integral \emph{t}  here requires  the actual \emph{U} to be much smaller than the previously predicted 0.7~eV for  EtMe$_3$Sb[Pd(dmit)$_2$]$_2$ \cite{calc3}.  In fact,  this  is naturally expected when strong electron-phonon interaction is considered,  polarization of the lattice would significantly reduce the effective on-site Coulomb repulsions. Assuming $U\sim$ 8~$t$ \cite{Theory1},  $U$ is probably in the order of 50 meV, which explains the rather accurate calculated energy levels, since the small differences between inter-orbital  and intra-orbital Coulomb interactions will not give much deviation.  Consequently, the upper Hubbard band should not be  far away from the LHB in EtMe$_3$Sb[Pd(dmit)$_2$]$_2$, which naturally explains the low excitation energy of about 40~meV estimated from the resistivity \cite{TransQSL} and the low charging onset temperature of 40~K in our photoemission study.  In other words, our results do suggest that EtMe$_3$Sb[Pd(dmit)$_2$]$_2$ is on the verge of a metal-insulator transition, in agreement with the theory \cite{Theory1, Theory3}. Consistently for $\kappa$-(ET)$_2$Cu$_2$(CN)$_3$, which is another  organic QSL  on a triangular lattice, the charge gap  is very small in optical conductivity measurements  (the exact number is hard to extract, probably in the 15$\sim$100~meV range) \cite{Chargegap1}, indicating the electronic structure characters found here might be fairly generic.


\begin{figure}[t]
\includegraphics[width=8.6cm]{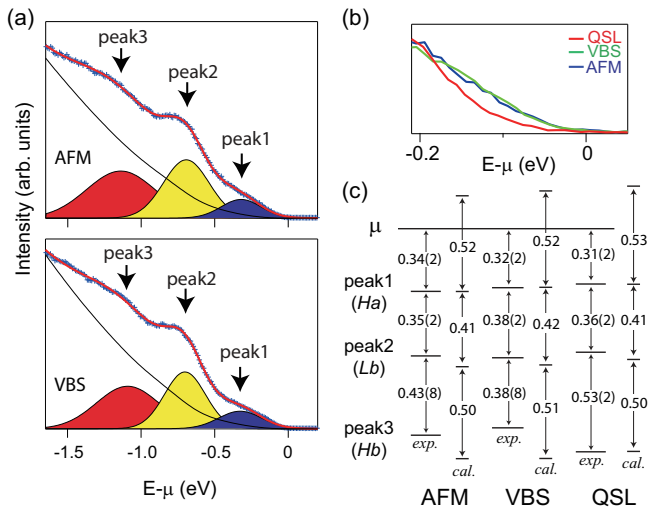}\caption{(color online)
(a) The normal-emission spectra of  Et$_2$Me$_2$P[Pd(dmit)$_2$]$_2$ (AFM compound) and EtMe$_3$P[Pd(dmit)$_2$]$_2$ (VBS compound)  are fitted with 3 Gaussian peaks plus a Shirley-type background, respectively. Data were taken with \emph{p} polarized 8~eV photons. (b) The $\Gamma$-point spectra of the three Pd(dmit)$_2$ salts in the vicinity of the chemical potential. (c) The schematic diagram for the experimental (\emph{exp.}) and calculated (\emph{cal.}) energy-levels of these materials. The calculated \emph{Ha} bands are individually aligned to the experimental positions of their peak1's. The unit is eV.}
\label{compare}
\end{figure}


For comparison, the electronic structure of Et$_2$Me$_2$P[Pd(dmit)$_2$]$_2$ (AFM) and EtMe$_3$P[Pd(dmit)$_2$]$_2$ (VBS) are  investigated, whose ARPES spectra with negligible dispersion are qualitatively similar to those of EtMe$_3$Sb[Pd(dmit)$_2$]$_2$. The typical spectra for these compounds in Fig.~\ref{compare}(a) demonstrate the insulating behavior of these Pd(dmit)$_2$ salts, while the ARPES spectral features in AFM and VBS materials are broader than those of QSL, and their energy separations are displayed in Fig.~\ref{compare}(c). The calculated energy levels of these Pd(dmit)$_2$ salts are also presented in Fig.~\ref{compare}(c), which show reasonable agreement with the experiments. Near the chemical potential, the spectral weight of the QSL is further suppressed from the chemical potential than the other two, as clearly illustrated in Fig.~\ref{compare}(b). For the QSL compound, peak1 is located at 0.31~$\pm$~0.02~eV with narrower width than those of VBS and AFM compounds, making the DOS of EtMe$_3$Sb[Pd(dmit)$_2$]$_2$ more suppressed near the chemical potential. This explains the relatively lower excitation energies of AFM ($\sim$~26 meV) and VBS ($\sim$~27 meV) compounds than that of QSL compound, as estimated from their resistivity data \cite{Trans1, TransVBS, TransQSL}. Our ARPES data could not provide information on the evolution of  $t$, $t'$, and $U$ in these compounds, because of their small amplitudes. However, we do observe certain variations in the energy level positions and spectral widths, which suggest the variations in the dimer structure and electron-boson interaction strength, respectively. Whether and how these tip the balance between different ground states require further investigation.

To summarize, we have presented  the  experimental electronic structure of  EtMe$_3$Sb[Pd(dmit)$_2$]$_2$, a QSL, for the first time, which naturally explains many phenomena observed in this compound. Moreover, we found that Pd(dmit)$_2$ salts in the QSL, VBS, and AFM states are Mott insulators with small gap and vanishing spectral weight at the chemical potential. Contrary to previous theoretical picture, we found that these systems are characterized by negligible hopping and broad polaronic lineshapes.  It  indicates that  strong electron-phonon/magnon interactions have to be included when describing these systems.  Our results suggest that a Holstein-Hubbard model with strong electron-boson coupling, instead of the Hubbard model, is more appropriate for a realistic modeling of the organic QSL's and related compounds.

\noindent\textit{Acknowledgements:} We gratefully acknowledge the helpful discussions with Prof. D. H. Lee and F. Wang. This work is supported in part by the National Science Foundation of China, National Basic Research Program of China (973 Program 2012CB921400), and Grant-in-aid for Scientific Research on Innovative Areas (No. 20110003) from the Ministry of Education, Culture, Sports, Science and Technology (MEXT), Japan.

\end{document}